\documentclass[aps,amsfonts,prl,twocolumn,superscriptaddress,showpacs]{revtex4}
\usepackage{epsfig,amsmath,amssymb,bm,epsf,graphics,psfrag,verbatim,subfigure}
\def\Prob{\mathcal{P}}

\usepackage[usenames]{color}

\newcommand{\bv}{\bm{{\rm b}}}

\newcommand{\qv}{\bm{{\rm q}}}

\newcommand{\kappat}{{\tilde \kappa}}
\newcommand{\rt}{{\tilde \kappa}}
\newcommand{\Tr}{{\rm Tr}}
\newcommand{\Remove}[1]{{}}
\newcommand{\TCLchange}[1]{{#1}}
\newcommand{\XMchange}[1]{{#1}}

\def\Prob{\mathcal{P}}

\def\fcNNN{\kappa}
\def\Tens{\mathbf}

\def\DyMa{D}
\def\GF{G}

\def\freq{\omega}

\def\fcNNNm{\kappa_{m}}

\def\VPert{V}

\def\bv{\hat{b}}
\def\T{T}

\def\freqst{\freq^{*}}

\begin{document}
\title{Soft modes and elasticity of nearly isostatic lattices: randomness and dissipation}
\author{Xiaoming Mao}
\affiliation{Department of Physics and Astronomy, University of
Pennsylvania, Philadelphia, PA 19104, USA }
\author{Ning Xu}
\affiliation{Department of Physics and Astronomy, University of
Pennsylvania, Philadelphia, PA 19104, USA } \affiliation{The James
Frank Institute, University of Chicago, Chicago, Illinois 60637}
\affiliation{Department of Physics, The Chinese University of Hong
Kong, Shatin, New Territories, Hong Kong}
\author{T. C. Lubensky}
\affiliation{Department of Physics and Astronomy, University of
Pennsylvania, Philadelphia, PA 19104, USA }

\date{\today}

\begin{abstract}
The square lattice with nearest neighbor central-force springs is
isostatic and does not support shear.  Using the Coherent Potential
Approximation (CPA), we study how the random addition, with
probability $\Prob=(z-4)/4$ ($z$ = average number of contacts), 
of next-nearest-neighbor ($NNN$) springs restores
rigidity and affects phonon structure. The CPA effective $NNN$
spring constant $\kappat_m(\omega)$, equivalent to the complex shear
modulus $G(\omega)$, obeys the scaling relation, $\kappat_m(\omega)
= \kappa_m h(\omega/\omega^*)$, at small $\Prob$, where $\kappa_m =
\kappat'_m(0)\sim \Prob^2$ and $\omega^* \sim \Prob$, implying
nonaffine elastic response at small $\Prob$ and the breakdown of
plane-wave states beyond the Ioffe-Regel limit at $\omega\approx
\omega^*$. We identify a divergent length $l^* \sim \Prob^{-1}$, and
we relate these results to jamming.
\end{abstract}


\maketitle

\TCLchange{Isostatic lattices
\cite{Maxwell1864,Alexander1998,Wyart2005} are systems at the onset
of mechanical stability in which the average number of contacts $z$
per particle in $d$-dimensions is equal to $z_c = 2d$.  A lattice
with $N$ particles and $N_c$ two-particle contacts has $N_0=dN-N_c$
zero modes.  An infinite isostatic lattice is one in which $N_c=N
z_c/2$, and the fraction of zero modes vanishes.  Because
particles at the boundary have fewer contacts than those in the
bulk, the number of zero modes in a finite isostatic lattice is
subextensive ($N_0\sim N^{(d-1)/d}$) and proportional to the
area of the system boundary.  As a result, the phonon spectrum of
isostatic lattices is one-dimensional in nature.} These properties
underly the elastic and vibrational properties of a variety of
systems including network glasses \cite{Phillips1981,Thorpe1983},
rigidity percolation \cite{JacobsThor1995,DuxburyMou1999},
$\beta$-cristobalite \cite{SwainsonDov1993a}, granular media~
\cite{EdwardsGrin1999,TkachenkoWit1999}, and networks
of semi-flexible polymers \cite{HeussingerFre2007}. Isostatic
lattices include $d$-dimensional hypercubic lattices and the $2d$
kagome, the $3d$ pyrochlore lattice, and their $d$-dimensional
generalizations \cite{Marck1998}, all with central-force springs
with spring constant $k$ connecting nearest neighbor ($NN$) sites.
They also include randomly packed spheres at the jamming transition
\cite{Durian1995,OhernNag2002,OhernNag2003}.

As in critical phenomena at ``standard" phase transitions, the
approach to the critical isostatic state, which this paper explores,
is characterized by diverging length and time scales and by scaling
behavior. Lattices can be moved off isostaticity in various ways,
including (1) introducing springs with a tunable spring constant
$\kappa$ connecting next nearest neighbor ($NNN$) sites
\cite{SouslovLub2009} and (2) increasing the volume fraction $\phi $
of packed spheres above the critical value $\phi_c$ at jamming
\cite{Durian1995,OhernNag2002,OhernNag2003,SilbertNag2005,WyartWit2005a,VitelliNag2009}.
The isostatic lattices with their soft modes are then approached
continuously as $\kappa$ or $\Delta \phi=(\phi - \phi_c)$ approach
zero, and divergent length scales $l^*$, vanishing frequencies
$\omega^*$, and possibly vanishing shear moduli $G$ (isotropic for
jamming and the anisotropic modulus $C_{44}\equiv C_{xyxy}$ for the
square lattice as detailed below) can be identified. In approach
(2), the number of contacts increases as $\Delta z = z-z_c \sim
(\Delta \phi)^{1/2}$, $l^* \sim (\Delta z)^{-1}$, $\omega^* \sim
\Delta z$, and $G\sim \Delta z$, whereas in approach (1) for the
square lattice $l^* \sim \kappa^{-1/2}$, $\omega^* \sim
\kappa^{1/2}$, and $G \sim \kappa$.

\begin{figure}
\centerline{\includegraphics{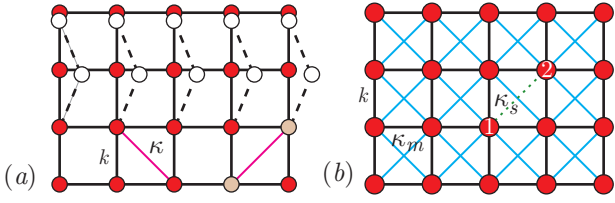}}
\caption{(a) Square lattice with $NN$ bonds with springs of spring
constant $k$ and $NNN$ bonds with randomly placed springs with
spring constant $\kappa$.  The distortion depicted with dotted lines
represents one of the zero modes of the lattice with no $NNN$
springs. (b) Effective-medium lattice with springs of spring
constant $\kappa_m$ on all $NNN$ bonds. In the CPA, the spring
constant $\kappa_s$ of a single $NNN$ bond is changed to $\kappa$ or
to $0$ with respective probabilities $\Prob$ and $1-\Prob$.}
\label{fig:sq-lattice}
\end{figure}

In this paper, we investigate a third approach to isostaticity in
the square lattice: we populate $NNN$ bonds with springs of spring
constant $\kappa$ with probability $\Prob$ as shown in
Fig.~\ref{fig:sq-lattice}.  At nonzero $\Prob$, \XMchange{the addition of an extensive number of $NNN$ bonds removes all zero modes with a probability that approaches unity~\cite{finite-modulus} as the number of sites $N\!\to\!\infty$, and as a result, the infinite lattice has a nonzero shear modulus for all $\Prob>0$.} 
Thus, our model describes a
rigidity percolation problem in which the percolation threshold is
at $\Prob \!= 0$. It is the particular case
\cite{Obukhov1995,MoukarzelLea1995} of the more general rigidity
percolation problem on a square lattice \cite{GarbocziThor1985} with
$NN$ and $NNN$ bonds populated independently with respective
probabilities $\Prob_{NN}$ and $\Prob$ in which $\Prob_{NN} \!=\! 1$.
This model shares underlying periodicity with approach (1) but it
includes randomness analogous to approach (2).  Adding a $NNN$
spring increases the number of contacts by $1$ so that 
$\Prob\!=(z-z_c)/4$, where $z_c=4$ in the $NN$ square lattice.  Unless
otherwise stated in what follows, we use reduced units with $k=1$ and
lattice constant $a=1$ and unitless spring constants, elastic
moduli, and frequencies: $\kappa/k \!\rightarrow\!\kappa$,
$G a^2/k \to G$, and
$\omega/\sqrt{k} \rightarrow \omega$.

\begin{figure}
\centerline{\includegraphics{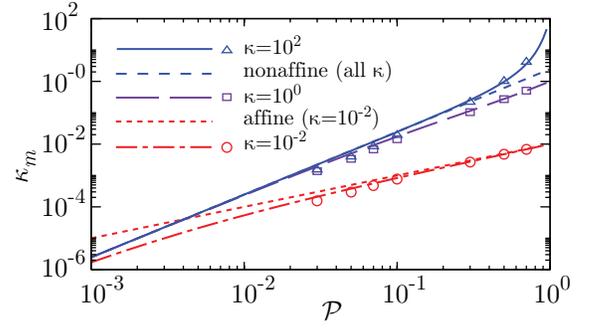}} \caption{(color
online) Comparison of the CPA solution (lines) and numerical
simulations on a $100\times 100$ lattice (data points) for the
effective medium spring constant $\fcNNNm$ as a function of $\Prob$
for $\kappa= 10^{-2}, 10^0$, and $10^2$ (in reduced units). Also
shown are the nonaffine ($\kappa_m = (\pi \Prob/2)^2$) and affine
limits ($\kappa_m = \Prob\kappa$). For the CPA at large $\Prob$, we
used the full dynamical matrix [Eq.~(\ref{eq:tensD-comp})] rather
than the approximate forms of Eq.~(\ref{eq:Dxx-low}).}
\label{fig:kappa-p}
\end{figure}

We study this random $NNN$ model using the Coherent Potential
Approximation (CPA) \cite{Soven1969,GarbocziThor1985,DasMac2007},
which gives good results for the conductivity of random networks
near percolation \cite{Kirkpatrick1973} and for rigidity percolation
problems \cite{GarbocziThor1985} except right in the vicinity of
$\Prob = \Prob_c$, and we verify that it gives results that are in
quantitative agreement with numerical simulations in our system.
\XMchange{ In the CPA, an effective medium of a uniform lattice with
every $NNN$ bond occupied by a spring with complex effective spring
constant $\kappat_m (\omega)= \kappat_m' (\omega) -
i\kappat_m^{\prime\prime}(\omega)$, determined by a proper
self-consistency condition, is used to capture the disorder average
of the random lattice.  
From $\kappat_m(\omega)$, which is also equal to the complex shear modulus $G(\freq)$, we can calculate (following the procedures of
approach (1) \cite{SouslovLub2009}) the characteristic length $l^*$
and frequency $\omega^*$ and the zero-frequency shear modulus
$G=\kappat'(\omega=0)$, as summarized in Table \ref{table}. As in
the case of jamming, $l^*\!\sim\!1/\freqst\!\sim\!(\Delta z)^{-1}$, in
agreement with the general cutting arguments of
Ref.~\cite{Wyart2005,WyartWit2005a}. The length $l^*$, being the
average distance between $NNN$ bonds in any row or column in the
random lattice, marks the crossover from $1d$ to $2d$ behavior in
the effective medium, because $NNN$ bonds couple neighboring $1d$
rows or columns.} The shear modulus, however, scales as
$G\!\sim\!\Prob^2\!\sim\!(\Delta z)^2$, rather than as $G\sim (\Delta z)$ at
jamming, implying highly nonaffine response near $\Prob = 0$.  If
the response were affine, every equivalent $NNN$ bond would distort
the same way in response to shear, and $G$ would be equal to $\Prob
\kappa$. Response becomes more nearly affine with $G\approx
\Prob\kappa$ when $\pi^2 \Prob \gg \kappa$. Figure \ref{fig:kappa-p}
shows $\kappa_m=G$ as a function of $\Prob$ for different
$\kappa$ calculated from the CPA and via numerical simulations using the conjugate gradient method~\cite{Press1986} to calculate
the relaxed response of the system to an applied shear.

\begin{table}
\caption{\label{table}Dependence of $l^*$, $\omega^*$, and $G$ on
$\Prob$ and $\Delta z$.}
\begin{ruledtabular}
\begin{tabular}{ccc|}
$l^*\sim \Prob^{-1}\sim(\Delta z)^{-1}$& $\omega^*\sim \Prob\sim
\Delta z$ & $G\sim \Prob^2 \sim(\Delta z)^2$
\end{tabular}
\end{ruledtabular}
\end{table}

\begin{figure}
\centerline{\includegraphics{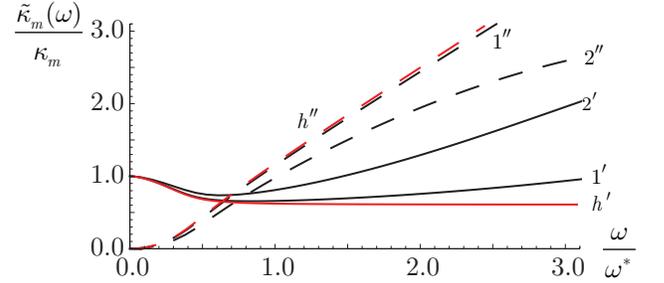}}
\caption{(color online) Real and imaginary parts of
$h(\omega/\omega^*)\equiv h' - i h''$ (labeled respectively $h'$ and
$h''$) and of $\kappa_m(\omega)/\kappa_m$ for $\Prob = 10^{-2}$ and
$10^{-1}$ (labeled respectively $1'$, $1''$, $2'$ and $2''$)for
$\kappa = 1$. Curves for $\Prob=10^{-3}$ and $10^{-4}$ differ by
less than $1\%$ from the $h$ curve and are not shown. The the full
dynamical matrix [Eq.~(\ref{eq:tensD-comp})] was used in the
$\Prob=10^{-1}$ calculation.}
\label{fig:kappat}
\end{figure}

The frequency dependence of $\kappat_m(\omega)$ is plotted in
Fig.~\ref{fig:kappat}. In the nonaffine regime, it obeys a scaling
law, $\kappat(\omega)= \kappa_m h(\omega/\omega^*)$, where $h(w)$
approaches unity as $w\rightarrow 0$. $\kappat''(\omega)$ vanishes
as $\omega^2$ at small $\omega$ but becomes nearly linear in
$\omega$ for $\omega \gtrsim 0.5 \omega^*$. This behavior
corresponds to a shear viscosity that vanishes as $\omega$ at small
$\omega$ but becomes a constant at large $\omega$.
A transverse phonon of frequency $\omega$ propagating along the
$y$-direction (i.e., with $q_x = 0$) has a wave number $q(\omega) =
\omega/\sqrt{\kappat_m'(\omega)}$ and a mean-free path $l(\omega) =
\sqrt{\kappa_m'(\omega)} \tau(\omega)$, where $\tau(\omega) =
2[\kappat_m''(\omega) q^2(\omega)/\omega]^{-1}$ is the decay time,
implying that the Ioffe-Regel limit \cite{IoffeReg1960} $q(\omega)
l(\omega) = 1$ occurs at $2\kappat_m'(\omega) =
\kappat_m''(\omega)$, i.e., at $\omega \approx \omega^*$.  Thus
$\omega^*$ sets the frequency scale for the nearly isostatic modes
and the scale at which plane-wave states become ill defined in
agreement with recent studies of thermal conductivity near jamming
\cite{VitelliNag2009}. Because $q_y(\omega^*) \sim\pi/a$, plane wave
states with $q_x=0$ are well-defined up to the zone edge.

\begin{figure}
\centerline{\includegraphics{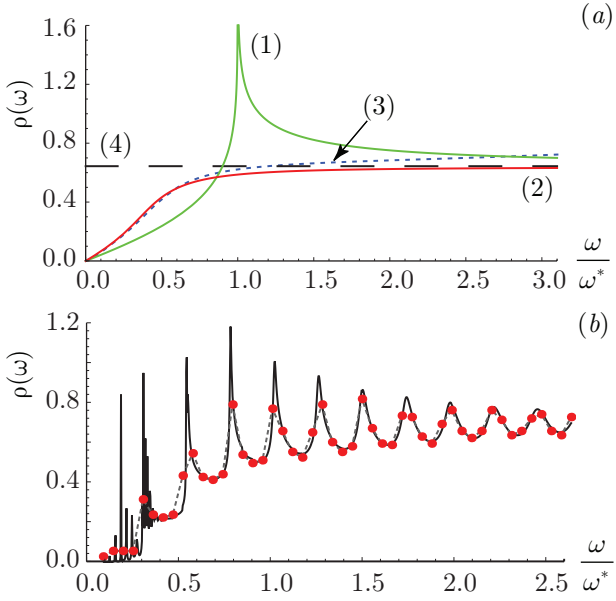}}
\caption{(color online) (a) Density of states $\rho(\omega)$ for (1)
(green) a uniform lattice with $\kappa = \kappa_m$ on all $NNN$
bonds, (2) (red) in the scaling nonaffine limit where
$\kappat_m(\omega) = \kappa_m h(\omega/\omega^*)$, and (3) (Blue
dotted) for $\Prob=10^{-1}$ (4) (black dashed) Isostatic 2-mode
limit of $2/\pi \approx 0.64$. (b) Density of states for a
$100\times100 $ lattice with $\Prob = 10^{-1}$ obtained via direct
numerical calculation (dots) and via the CPA (line) using the full
rather than the approximate dynamical matrix of
Eq.~(\ref{eq:Dxx-low}). Binning of the CPA result would wash out the
spikes at low frequency at $\omega = q_x = 2 \pi n /100$ (for
integer $n$).}
\label{fig:dos}
\end{figure}

\TCLchange{Because the zero modes on isostatic square lattice are uniform displacements of rows or columns, its phonon spectrum is identical to that of decoupled one-dimensional chains
with frequencies $\omega_{x,y} ( \qv ) = 2 |\sin q_{x,y}/2|$ and
density of states $\rho(\omega) = (2/\pi)/\sqrt{4  - \omega^2}$ with
a nonzero value $1/\pi$ at $\omega = 0$ as shown in
Fig.~\ref{fig:dos}.  When the effective-medium $NNN$} coupling
$\kappat_m(\omega)$ is added, the dynamical matrix becomes
\begin{align}
& D_{xx} ( \qv )  \!=\! D_{yy}(q_y,q_x)=  4 \sin^2 (q_x /2)\!+\! 4 \kappat_m(\omega) \sin^2 (q_y /2) \nonumber\\
& + 4 \kappat_m(\omega) \sin^2 (q_x /2) - 8 \kappat_m(\omega) \sin^2 (q_x /2)\sin^2 (q_y /2) , \nonumber  \\
& D_{xy} ( \qv)  = D_{yx}(\qv)= 2 \kappat_m(\omega) \sin (q_x ) \sin
(q_y ) .
\label{eq:tensD-comp}
\end{align}
In the $\qv\rightarrow 0$ limit, the dynamical matrix reduces to that of
continuum elastic theory with $D_{xx} = C_{11} q_x^2 + C_{44}
q_y^2$, where $C_{11}$ is a compression modulus and $C_{44}$ the
shear modulus.  $C_{44} ( \omega)$ is the complex shear storage
modulus $G(\omega)$. Comparison of the continuum form with the small
$\qv$ limit of Eq.~(\ref{eq:tensD-comp}) yields $\kappat_m(\omega) =
G(\omega)$.

When $|\kappat_m ( \omega)| \ll 1$, the off-diagonal terms in
$D_{ij}$ can be ignored, and the low-frequency modes follow
from
\begin{equation}
D_{xx}(\qv) \approx  q_x^2 + 4 \kappat_m(\omega) \sin^2 (q_y /2) \approx
q_x^2 + \kappat_m ( \omega) q_y^2
\label{eq:Dxx-low}
\end{equation}
and a similar approximation for $D_{yy}(\qv)$.  
Replacing $\kappat_m(\omega)$ by its $\freq\to 0$ limit $\kappa_m$ 
yields a characteristic length $l^* =\sqrt{1/4 \kappa_m }$ through
the comparison of $q_x^2$ with $D_{xx} (0,\pi) = 4 \kappa_m$ and
a characteristic frequency at the zone edge of $\omega^* =
\sqrt{D_{xx} ( 0 , \pi)} = 2 \sqrt{\kappa_m}$. For $q_x\! > \!1/l^*$ (or
$\omega\!>\!\omega^*$), the excitation spectrum is one-dimensional in $q_x$.
These observations along with $\kappa_m \sim \Prob^2$, which we
derive below, lead to the results of Table \ref{table}.

To proceed with the CPA, we use the $2 \times 2$ phonon matrix
Green's function of this effective medium
\begin{eqnarray}
\Tens{\GF}(\qv,\freq)=[\freq^2 \Tens{I}-\Tens{\DyMa}(\qv)]^{-1}.
\end{eqnarray}
In the CPA approximation \cite{Soven1969,Kirkpatrick1973}, an
arbitrary NNN bond, say, between particles 1 and 2 as shown in
Fig.~\ref{fig:sq-lattice}(b), is replaced by a new one with a random
spring constant $\kappa_s$ with values $\kappa$ and $0$ with
respective probabilities $\Prob$ and $1-\Prob$. The dynamical matrix
then changes to $\Tens{\DyMa}^V = \Tens{\DyMa}+ \Tens{\VPert}$,
where $\Tens{\VPert}$ is the potential given by
\cite{GarbocziThor1985}
\begin{eqnarray}\label{EQ:VReal}
\Tens{\VPert}_{l,l'} \!=\! (\kappa_s\!-\!\kappat_m)
(\delta_{l,1}-\delta_{l,2})\bv \otimes
(\delta_{l',1}-\delta_{l',2})\bv ,
\end{eqnarray}
in real space, $\bv=(\mathbf{e}_x+\mathbf{e}_y)/\sqrt{2}$ is the
unit vector along the chosen NNN bond, and $l$ and $l'$ specify
sites on the lattice. The potential $\Tens{\VPert}$ leads to a
modification of the phonon Green's function, $G_{l,l'}^V
(\omega)$, which can be calculated following standard procedures:
\begin{eqnarray}
\Tens{\GF}_{l,l'}^V(\freq) = \Tens{\GF}_{l-l'}(\freq)\!+\!
\sum_{l_1,l_2} \Tens{\GF}_{l-l_1}(\freq)\! \cdot \Tens{\T}_{l_1,l_2}
\! \cdot \Tens{\GF}_{l_2-l'}(\freq),
\end{eqnarray}
where $\Tens{G}_{l-l'}$ is the Fourier transform with respect to
$\qv$ of $\Tens{G}(\qv,\omega)$ and where $\Tens{\T}=[\Tens{1} -
\Tens{\VPert}\cdot \Tens{\GF}]^{-1}\cdot \Tens{\VPert}$ is the
scattering $T$-matrix. The effective spring constant
$\kappat_m(\omega)$ is determined within the CPA through the
requirement that the average $\Tens{\T}$ vanish: $\Prob \,
\Tens{\T}\vert_{\kappa_s = \kappa} + (1-\Prob)\,
\Tens{\T}\vert_{\kappa_s=0} =0$ so that
\begin{equation}
f(\rt_m, \omega) \rt_m^2(\omega) - [1+\kappa f(\rt_m,
\omega)]\rt_m(\omega) + \kappa \Prob = 0
\label{eq:CPAeq} .
\end{equation}
The function $f$ can be expressed as $f(\rt_m,\omega) =[ 2/(\pi
\sqrt{\rt_m})] {\tilde g}(\rt_m,\omega/\sqrt{\rt_m})$, where
\begin{equation}
{\tilde g}(r,s) = \frac{1}{2}\int_0^{\pi} dq
\frac{1-e^{-\sqrt{r}p(q,s)}\cos q}{p(q,s)} ,
\label{eq:CPAint}
\end{equation}
with $p(q,s) = \sqrt{4 \sin^2(q/2) - s^2}$.  In the limit $r,
s\rightarrow 0$, ${\tilde g}(r,s) = 1$, and thus $f(\kappat_m, 0 )
\rightarrow [2/(\pi \sqrt{\kappa_m})]$ as $\kappa_m \rightarrow 0$.
When $\sqrt{r} p(\pi, s) \ll 1$, the exponential in the numerator of
${\tilde g}(r,s)$ can be replaced by unity, and ${\tilde g}(0,s)
\equiv g(s)$, $g(s) \rightarrow 1+
(s^2/8)\{\ln[8/(\sqrt{e}s)]+i(\pi/2)\}$. We expect $\kappa_m$ to
tend to zero with $\Prob$ so that in the small $\Prob$ limit, we can
generally ignore the first term in Eq.~(\ref{eq:CPAeq}).

We consider first the static limit, $\freq=0$, for which the
self-consistency equation for small $\Prob$ becomes
\begin{equation}\label{EQ:QuadS}
    \fcNNNm+\frac{2\fcNNN}{\pi}\sqrt{\fcNNNm}-\Prob\fcNNN=0.
\end{equation}
The solution of this equation has two limits:
\begin{equation}
\fcNNNm \simeq
\begin{cases}
\big(\pi\Prob/2\big)^2  & \text{if
    $\pi^2\Prob\ll \fcNNN$,}
    \\
   \Prob\fcNNN & \text{if $\pi^2\Prob\gg \fcNNN$,}
\end{cases}
\end{equation}
\XMchange{as shown in Fig.~\ref{fig:kappa-p}, together with
solutions of the full CPA equation~(\ref{eq:CPAeq}) and numerical
simulations.  } In the first case, $\kappa
\sqrt{\kappa_m}\gg\kappa_m$, and the solution for $\kappa_m$ is
obtained by ignoring the first term in Eq.~(\ref{EQ:QuadS}); in the
second case, the opposite is true, and $\kappa_m$ is obtained by
ignoring the second term in this equation. In the second case, every
$NNN$ bond distorts in the same way under stress, and response is
affine.  In the first case $\kappa _m = (\pi \Prob/2)^2 \ll \Prob
\kappa$, and response is nonaffine with local rearrangements in
response to stress that lower the shear modulus to below its affine
limit. Within the CPA, this result emerges because of the divergent
elastic response encoded in $\Tens{G}$ (and $f(\kappa_m,0)$) as
$\kappa_m \rightarrow 0$.
As $\kappa$ approaches zero at fixed $\Prob$, distortions produced
by the extra bond decrease and the nonaffine regime becomes
vanishingly small.

\XMchange{ For finite frequency $\freq$, the effective medium spring
constant is complex, $\kappat(\freq)=\kappat'(\omega)
-i\kappat''(\omega)$, where the imaginary part $\kappat''(\omega)$,
which is odd in $\freq$ and positive for $\freq>0$, describes
damping of phonons in this random network. As in the static case,
the nonaffine limit of the CPA result for $\kappat ( \freq)$ at
small $\Prob$ is the solution to $\rt_m f(\rt_m, \omega) = \Prob$
obtained from Eq.~(\ref{eq:CPAeq}) by ignoring all but its last two
terms. Following Eq.~(\ref{eq:CPAint}), at small $\rt_m$ and
$\omega$, $f(\rt_m, \omega) = [2/(\pi
\sqrt{\rt_m})]g(2\sqrt{\kappa_m/\rt_m}\,\omega/\omega^*)$. Thus in
this limit, $\kappat_m(\omega)$ satisfies a scaling equation
$\kappat_m(\omega) = \kappa_m h(\omega/\omega^*)$.  As $\omega
\rightarrow 0$, $h(w) \rightarrow 1 -w^2 \{\ln [4/(\sqrt{e}
w)]+i(\pi/2)\}$, and $\kappat''(\omega) \sim \omega^2$ at small
$\omega$. We calculated $\kappat_m(\omega)/\kappa_m$ for $\Prob =
10^{-4}, 10^{-3}, 10^{-2}$ and $10^{-1}$ with the full CPA
equation~(\ref{eq:CPAeq}) and the nonaffine scaling function
$h(\omega/\omega^*)$ for $\kappa = 1$.
The crossover from nonaffine to affine behavior in the static limit
is at $\Prob = 1/\pi^2 \approx 10^{-1}$, so all cases but $\Prob =
10^{-1}$ are at or near the nonaffine limit.
$\kappa_m''(\omega)$ becomes greater than $\kappa_m'(\omega)$, and
thus according the Ioffe-Regel criterion \cite{IoffeReg1960},
plane-wave phonon modes become heavily damped and ill-defined at
$\omega \approx \omega^*$ for all four values of $\Prob$. }

\TCLchange{The phonon density of states (DOS) $\rho(\omega)$,
calculated from $\textrm{Im}\Tr \Tens{G}^m ( \qv, \omega)$ in the
usual way, is plotted in Fig.~\ref{fig:dos}(a) as a function of
$\omega/\omega^*$. Curves for the three lowest $\Prob$ in
Fig.~\ref{fig:dos}(a) collapse on to a common curve for $\omega \leq
3\omega^*$.  The curve for $\Prob = 10^{-1}$ departs from the common
curve at $\omega\approx 0.5 \omega^*$ and is plotted in the figure.
The large value of $\kappa''_m(\omega^*)$ in the random system
removes the strong van Hove singularity at $\omega^*$ of the uniform
system.  Figure \ref{fig:dos}(b) compares the DOS for a finite
lattice calculated from CPA and by direct numerical 
diagonalization of the Hessian matrix using ARPACK~\cite{ARPACK}.
The peaks in Figure \ref{fig:dos}(b) at $\omega = q_x = (2\pi n/L)$
are due to finite size effects of the lattice with size $L$.}

\TCLchange{We have used the CPA to analyze the static and dynamic
properties of a simple system on the threshold of isostaticity,
namely a square lattice with $NN$ springs and randomly distributed
$NNN$ springs.  This system provides clean analytic results
about a random system near isostaticity, including nonaffine
response near $\Prob=0$, and the scaling form for
$\kappat_m(\omega)$ (which to our knowledge has not been observed in
jamming systems), that can serve as a comparison point for more
complicated systems. Our results strongly suggest that the divergent
length $l^*\sim 1/\omega^*\sim(\Delta z)^{-1}$ is a common feature
of all nearly isostatic systems in agreement with the arguments of
Ref.~\cite{Wyart2005}.  They also unambiguously demonstrate that
elastic moduli are not universal but depend on the geometry of the
isostatic lattice.  Further study is needed to determine exactly
what properties of the isostatic lattice lead for example to a
finite bulk modulus and a shear modulus vanishing as $\Delta z$ (as
in jamming) or $(\Delta z)^2$ (current system) or as $(\Delta z)^0$
(kagome lattice \cite{MaoLub2010}) or to one in which both $B$ and
$G$ vanish as $\Delta z$ as in Ref. \cite{EllenbroekHec2009}} .

We are grateful for helpful discussions with Andrea Liu and Anton
Souslov. This work is supported in part by NSF-DMR-0804900.


\begin{thebibliography}{26}
\bibitem{finite-modulus}
\bibfnamefont{Our MC simulations on systems up to $160 \times 160$ show that the probability $\Prob_t$ at which a finite fraction of random configurations are rigid vanishes a $L^{-\beta}$ where $\beta\sim 0.75$ and thus that rigidity percolation occurs at $\Prob=0$.}
\end{thebibliography}

\end{document}